\documentstyle[amssymb,prb,aps]{revtex}

\begin{document}
\title{Anelastic spectroscopy study of the spin-glass and cluster spin-glass phases
of La$_{2-x}$Sr$_{x}$CuO$_{4}$ $(0.015<x<0.03)$}
\author{A. Paolone$^{1}$, F. Cordero$^{2}$, R. Cantelli$^{1}$, M. Ferretti$^{3}$}
\address{$^{1}$ Universit\`{a} di Roma ``La Sapienza``, Dipartimento di Fisica, P.le\\
A. Moro 2, I-00185 Roma, and INFM, Italy}
\address{$^{2}$ CNR, Area di Ricerca di Tor Vergata, Istituto di Acustica ``O.M.\\
Corbino``, Via del Fosso del Cavaliere 100, I-00133 Roma, and INFM, Italy}
\address{$^{3}$ Universit\`{a} di Genova, Dipartimento di Chimica e Chimica Fisica,\\
Via Dodecanneso 31, I-16146 Genova, and INFM, Italy}
\maketitle

\begin{abstract}
The anelastic spectra of La$_{2-x}$Sr$_{x}$CuO$_{4}$ have been measured at
liquid He temperatures slightly below and above the concentration $%
x_{c}\simeq 0.02$ which is considered to separate the spin-glass phase from
the cluster spin-glass (CSG) phase. For $x\le x_{c}$ all the elastic energy
loss functions show a step below the temperature $T_{g}\left( x=0.02\right) $
of freezing into the CSG state, similarly to what found in samples well
within the CSG phase, but with a smaller amplitude. The excess dissipation
in the CSG state is attributed to the motion of the domain walls between the
clusters of antiferromagnetically correlated spin. These results are in
agreement with the recent proposal, based on inelastic neutron scattering,
of an electronic phase separation between regions with $x\sim 0$ and $x\sim
0.02$, at least for $x>0.015$.
\end{abstract}


\twocolumn

\section{INTRODUCTION}

The magnetic properties of the High $T_{c}$ Superconductors (HTCS) have
attracted much interest also because these compounds are experimental
realizations of a 2D Heisenberg antiferromagnet.\cite{RBC98} In undoped La$%
_{2}$CuO$_{4}$ the Cu$^{2+}$ spins order into a 3D antiferromagnetic (AF)
state below $T_{N}\simeq 315~$K with the staggered magnetization in the $ab$
plane.\cite{VSM87} When doping by substituting Sr for La, the long range AF
order is rapidly suppressed and around $x_{c}\simeq 0.02$ the N\'{e}el
temperature drops to 0~K. In the long range AF region $T_{\text{N}}\left(
x\right) $ follows a power law relationship with $x$, interpreted as an
indication that the holes introduced by doping form walls separating domains
of undoped material.\cite{CCJ93} Later work indicated that the spin degrees
of freedom associated with the doped holes are distinct from the in-plane Cu$%
^{2+}$ spin degrees of freedom that order themselves below $T_{\text{N}}$,
and the localization of the doped holes allows the associated spins to
progressively slow down and freeze.\cite{RBC98,CBC93} The state in which the
doped spins freeze is usually referred to as a spin glass (SG) state, and
occurs below $T_{f}\left( x\right) \simeq \left( 815~\text{K}\right) x$. For 
$x>x_{c}$ there is no long range AF order, but approaching $T_{g}\simeq 0.2$%
~K$/x$ AF correlations develop within domains separated by charge walls,
with the easy axes of the staggered magnetization uncorrelated between
different clusters. This picture corresponds to a cluster spin-glass (CSG)
state. The formation of the SG and CSG states are inferred from sharp maxima
in the $^{139}$La NQR\cite{RBC98,CBJ92,JBC99} and $\mu $SR\cite{NBB98}
relaxation rates, which indicate the slowing of the AF fluctuations below
the measuring frequency ($\sim 10^{7}-10^{8}$~Hz in those experiments), and
from the observation of irreversibility, remnant magnetization, and scaling
behavior in magnetic susceptibility experiments.\cite{CBK95,WUE99} Also
anelastic spectroscopy can provide useful information about the magnetic
properties of high HTCS. Above $x_{c}$, the elastic energy loss coefficient
shows a rise below a temperature $T_{on}$ close to the $T_{g}\left( x\right) 
$ for freezing into the CSG state.\cite{CPC00,92} The absorption is not
peaked near $T_{g}$, but is step-like or at least displays a plateau, and
therefore does not directly correspond to the peak in the dynamic spin
susceptibility due to the spin freezing; rather, it has been attributed to
the stress-induced changes of the sizes of the spin clusters, or
equivalently to the motion of the domain walls.\cite{CPC00,92}

Recently, Matsuda {\it et al.} reported a neutron scattering study of the
magnetic correlations in La$_{2-x}$Sr$_{x}$CuO$_{4}$ for $x<x_{c}$, which
suggests a different picture of the spin glass phase.\cite{MFY01} In fact,
they found that also at $x<x_{c}$ the 3D AF ordered phase coexists below $%
\sim 30$ K with domains of ''diagonal'' stripe phase (with the hole stripes
at $45^{\text{o}}$ with respect to the Cu-O bonds), as observed for $x>x_{c}$%
. According to these authors, the hole localization starting around 150~K
involves an electronic phase separation into regions with $x_{1}\sim 0$ and $%
x_{2}\sim 0.02$. The volume fraction of the $x_{2}=0.02$ phase changes as a
function of the Sr doping, in order to achieve the average $x$.

In the following we will report on anelastic spectroscopy measurements of
lightly doped La$_{2-x}$Sr$_{x}$CuO$_{4}$, where the anelastic spectra for $%
x<x_{c}$ present the same features attributed to the domain wall motion into
the CSG phase, although attenuated. The step-like rise of dissipation occurs
near 10~K, as for $x\simeq 0.02$, supporting the view of an electronic phase
separation, at least for $0.015<x<0.02$.

\section{EXPERIMENTAL AND RESULTS}

The samples where prepared by standard solid state reaction as described in
Ref.~\onlinecite{DFF94} and cut in bars approximately $40\times 4\times 0.6$%
~mm$^{3}$. In the as-sintered state all the samples contained small amounts
of interstitial O, which was outgassed by heating in vacuum up to $790~$K.
The complex Young's modulus $E\left( \omega \right) =E^{\prime }+iE^{\prime
\prime }$, whose reciprocal is the elastic compliance $S=E^{-1}$, was
measured as a function of temperature by electrostatically exciting the
flexural modes. The vibration amplitude was detected by a frequency
modulation technique. The vibration frequency, $\omega /2\pi $, is
proportional to $\sqrt{E^{\prime }}$, while the elastic energy loss
coefficient (or reciprocal of the mechanical $Q$) is given by\cite{NB} $%
Q^{-1}\left( \omega ,T\right) =$ $E^{\prime \prime }/E^{\prime }=$ $%
S^{\prime \prime }/S^{\prime }$, and was measured by the decay of the free
oscillations or the width of the resonance peak. The imaginary part of the
dynamic susceptibility $S^{\prime \prime }$ is related to the spectral
density $J_{\varepsilon }\left( \omega ,T\right) =$ $\int dt\,e^{i\omega
t}\left\langle \varepsilon \left( t\right) \varepsilon \left( 0\right)
\right\rangle $ of the macroscopic strain $\varepsilon $ through the
fluctuation-dissipation theorem, $S^{\prime \prime }\propto $ $\left( \omega
/2k_{\text{B}}T\right) J_{\varepsilon }$. Magnetoelastic coupling can couple
strain to the spin degrees of freedom, so that pseudodiffusive spin
excitations (with null characteristic frequency) or the motion of magnetic
boundaries can contribute to $J_{\varepsilon }$. An elementary relaxation
process with a thermally activated relaxation time $\tau \left( T\right) $
contributes with $S^{\prime \prime }\propto $ $\omega \tau /\left[ 1+\left(
\omega \tau \right) ^{2}\right] $, peaked at the temperature at which $%
\omega \tau =1$ (which therefore increases with increasing $\omega $), while
the relaxation of extended and interacting structures like domain walls
generally gives rise to a broader dissipation curve.

The nominal compositions of the samples were: $x_{\text{nom}}=$ $0.015$,
0.016, 0.018, 0.024 and 0.030. The final Sr contents were checked from the
temperature position of the step in the Young's modulus $E$ due to the
tetragonal (HTT) / orthorhombic (LTO) transition, which occurs at a
temperature $T_{t}\left( x\right) $ decreasing with doping approximately as $%
T_{t}\left( x\right) =\left( 535-x/0.217\right) $~K (Ref.~\onlinecite{Joh97}%
). The shape of the anomaly in $E$ cannot be completely fitted by a simple
model, since it includes both contributions from the coupling of the
spontaneous strain with the soft mode\cite{SMM} and from the domain wall
motion,\cite{LLN90} but its position in temperature provides information
about the level of doping. Figure 1 presents the logarithmic derivative of
the Young's modulus with respect to temperature, $d\ln \left[ E/E\left(
0\right) \right] /dT$, so that the steps become peaks whose widths provide
upper limits to the spread of the local Sr concentration; the peaks are
fitted with lorentzians. The Sr concentrations estimated from the peak
positions and from $T_{t}\left( x\right) $ are $x=0.0165$, 0.0155, 0.0183,
0.0227 and 0.0273 respectively; the sample with $x_{\text{nom}}=0.016$
results to be less doped than the one with $x_{\text{nom}}=0.015$ and with a
broader transition. There is considerable overlapping of the peaks of the
samples with $x_{\text{nom}}<0.02$, but this should not be simply
interpreted as a distribution of local $x$ spanning 0.015-0.018 for all
these samples. In fact, the peaks in the derivative of $E\left( T\right) $
cannot be interpreted as the distribution function of the local $x$ after
the $T$ scale is converted into an $x$ scale inverting $T_{t}\left( x\right) 
$, since the step of $E\left( T\right) $ has an intrinsic width due to the
progressive phonon softening above $T_{t}$ and to the relaxational character
of the domain wall motion below $T_{t}$. The undoped samples, for example,
certainly do not have a spread of Sr concentration, but the widths of their
lorentzians (not shown here) are $16-22$~K, compared to 10~K of the sample
with $x_{\text{nom}}=0.016$. In the following we will refer to the doping
level obtained from the position of the step in the Young's modulus.

Figure 2 presents the anelastic spectra of the five samples measured
exciting the first flexural mode, around 1~kHz. The sharp rise of
dissipation at the lowest temperatures is the tail of an intense peak
attributed to the tunneling-driven tilt motion of a fraction of the O
octahedra.\cite{61,76} The shift of the peak to lower temperature with
increasing doping would be due to a direct coupling between the hole
excitations and the tunneling-driven local tilts of the O\ octahedra:\ the
more the hole excitations and the faster the local tilting.\cite{76} We are
concerned with the steplike feature at $T_{g}\left( x\right) \simeq 0.2/x$
for $x>0.02$, and below $\simeq 10$~K for $x<0.02$. That such a dissipation
rise occurs at $T_{g}$ and is steplike or at least consists of a plateau
rather than a peak is particularly evident at higher doping, when the
influence of the tail of the peak at lower temperature is less important;%
\cite{CPC00} here only two curves with $x\gtrsim 0.02$ are presented. The
dissipation rise has been attributed to the stress-induced movement of the
boundaries between the clusters of quasi-frozen antiferromagnetically
correlated spins which form the CSG phase.\cite{CPC00} The main result is
that also the samples with $x<0.02$ present a similar feature around 10~K.
With lowering $x$ below 0.02, the position of the dissipation step remains
unchanged, while its intensity decreases; this is more difficult to see for
the sample with $x=0.0155$, since the tail of the low temperature peak is
shifted to higher temperature (consistent with the lowest doping in spite of
the nominal $x$). Considering also previous data\cite{CPC00}, it appears
that, starting from high doping and lowering it, the dissipation step first
rises in temperature according to $T_{g}\left( x\right) $ and increases in
intensity down to $x\simeq 0.02$; below that doping its temperature remains
unchanged and the intensity decreases.

\section{DISCUSSION}

One would expect that the dissipation rise near 10~K for $x<0.02$
corresponds to the peak in the spin susceptibility occurring when the
magnetic fluctuations slow down below the measuring frequency during the
freezing process into the SG phase, as in the NQR measurements;\cite
{RBC98,CBJ92,JBC99} also the temperature region would be as expected, since,
according to the generally accepted phase diagram,\cite{Joh97} at $x=x_{c}$
both $T_{f}\left( x\right) $ and $T_{g}\left( x\right) $ merge at $10-15$~K.%
{\em \ }However, there are two features signalling the different nature of
the absorption rise below $\sim 10$~K: one is the shape, certainly different
from the sharp peak of the NQR relaxation rate, and the other is the total
absence of the expected linear shift toward lower temperature when the
doping is reduced. This is better demonstrated by the derivatives of the $%
Q^{-1}\left( T\right) $ curves in Fig. 3, where the negative peaks at high
temperature, labelled with the estimated doping values, correspond to the
steps in $Q^{-1}\left( T\right) $. The three peaks with $x<0.02$ even shift
to higher temperature with decreasing $x$, although the effect is very small
and most likely attributable to the increasing influence of the tail of the
peak at lower temperature. Instead, according to the generally adopted phase
diagram with $T_{f}\left( x\right) \simeq x\times \left( 815~\text{K}\right) 
$ the three peaks should display an overall shift by 2.3~K in the opposite
direction; it is therefore clear from Fig. 3 that there is no relation
between the acoustic dissipation step and $T_{f}\propto x$.

The spectra of the samples with $x<0.02$ strictly resemble those for $x>0.02$%
, and the latter have been successfully interpreted in terms of onset below $%
T_{g}$ of the motion of the domain walls between different spin clusters in
the CSG phase. The decrease of the amplitude $\Delta $ of the absorption
step above $x=0.02$ has been semiquantitatively explained considering a
simple model of the CSG phase, with the Sr atoms acting as pinning points
for the domain walls, which coincide with the hole stripes.\cite{CPC01} The
relaxation strength is expected to be of the form $\Delta \propto
n\,l^{\alpha }$, where $n$ is the volume concentration of the domain walls, $%
l$ is the distance between pinning points, and $\alpha $ turns out $\simeq
2.5$, intermediate between the case of the motion of dislocations and of
walls between ferromagnetic domains.\cite{CPC01} It was also possible to
observe the pinning of the walls by the low-temperature tetragonal lattice
modulation in samples doped with Ba instead of Sr.\cite{CPC01} The mechanism
of dissipative motion of the domain walls is well known for ferromagnetic
materials,\cite{NB} and is possible also for an AF state, if an anisotropic
strain is coupled with the easy magnetization axis. In this case, the
elastic energy of domains with different orientations of the staggered
magnetization are differently affected by a shear stress, and the lower
energy domains grow at the expenses of the higher energy ones providing the
coupling between strain and domain wall motion. The same mechanism could not
be applied to the SG state, even admitting the existence of hole stripes
below 10~K, since they would move into a uniform long-range ordered AF
matrix, instead of separating inequivalent domains, and other more subtle
mechanisms should be invoked.

It has to be concluded that the anelastic spectra for $x<0.02$ (Fig. 2a)
cannot be justified in terms of the spin-glass phase with $T_{f}\left(
x\right) $ increasing linearly with $x$; rather, they can be explained in a
straightforward way within the physical picture derived from the inelastic
neutron scattering measurements.\cite{MFY01} According to these
observations, there is an electronic phase separation below $x=0.02$, where
domains with fixed $x_{2}=0.02$ coexist with regions of undoped material
with $x_{1}=0$. Such domains have sizes estimated in several hundreds of
\AA\ within the CuO$_{2}$ planes, and display the same ''diagonal stripe''
correlations which are observed in the CSG state for $x\gtrsim 0.02$. The
volume fraction of these domains has also been verified to increase linearly
with $x$, as expected. In this phase separation picture, the elastic energy
dissipation curves for $x<0.02$ simply contain the step-like increase due to
the motion of the hole stripes within the $x_{2}\simeq 0.02$ domains, whose
sizes are sufficiently large that they appear as the homogeneous phase $%
x=0.02$ with the same $T_{g}\simeq 10$~K. The reduced volume fraction at
lower doping simply results in a reduced amplitude of the elastic energy
absorption.

The fact that the temperature at which our measurements reveal the presence
of spin domains is much lower than the temperature of 30~K reported by
Matsuda {\it et al.}\cite{MFY01} also agrees with what is observed in the
CSG state.\cite{MCP01} In fact, the onset temperature for freezing toward
the glass state, $T_{on}$, depends on the time scale at which the
experimental technique probes the system, and decreases as the angular
frequency $\omega $ of the probe decreases. In neutron scattering
experiments $\omega $ is of the order of $4\cdot 10^{11}$~s$^{-1}$, whilst
for the anelastic spectroscopy $\omega \sim 10^{4}\div 10^{5}$~s$^{-1}$,
resulting in a factor 2.5 between the two $T_{on}$ for at $x\sim 0.03$;\cite
{MCP01} this is consistent with the $T_{on}$ $=10\div 15$~K deduced from the
curves of Fig.~2a.

The dissipation curves of Fig. 2a clearly indicate that the same dissipation
mechanism of the CSG phase is operative also at $x<0.02$, and therefore are
in good agreement with the phase separation picture from the neutron
scattering experiment;\cite{MFY01} however, these curves alone are not
sufficient for clearly discriminating between a neat phase separation into $%
x_{1}\sim 0$ and $x_{2}\sim 0.02$ or a situation with a smooth transition
from the CSG to the SG state over a wide concentration range around 0.02.
For example, according to the theoretical model proposed by Gooding {\it et
al.,}\cite{GSB97} at low temperature the holes localize near the Sr dopants
and circulate over the four Cu atoms neighbors to Sr, inducing a distortion
of the surrounding Cu spins, otherwise aligned according to the prevalent AF
order parameter. The spin texture arising from the frustrated combination of
the spin distortions from the various localized holes produces domains with
differently oriented AF order parameters, which can be identified with the
frozen AF spin clusters. In this model there is no clear boundary between
the SG and CSG state, and the present data could also fit into such a
description.

The possibility has also to be considered that the discrepancy between the
present results below $x\simeq 0.02$ and the canonical phase diagram is due
to inhomogeneous doping of the samples at a microscopic level. Errors in
evaluating the actual doping and inhomogeneous doping would produce
particularly large shifts and uncertainty in the determination of $T_{f}$
and $T_{g}$, since these temperatures appear to merge around the value $%
10-15 $~K at $x=x_{c}=0.02$ , where they strongly depend on doping. Indeed,
there is a wide scattering of experimental data in that region of the phase
diagram,\cite{Joh97,EPJB} which can be due to uncertainties in $x$, but also
to the fact that the transition between SG and CSG is not as sharp as
supposed. The present data support the second explanation. In fact, we
cannot exclude the possibility of inhomogeneous doping of our samples with
certainty, only based on the width of the HTT/LTO phase transition, because
we are not able to carry out a rigorous analysis of the curves of Fig. 1.
Still, we note that if the high temperature of the step of samples with
average $x<0.02$ is due to regions with $x>0.02$, then we should expect a
similar spread of the local doping also for the samples with $x>0.02$, since
all the samples have been prepared in the same way. Then, the sample with $%
x=0.0273$ should exhibit a partial onset of dissipation already above 10~K,
due to regions with $x\simeq 0.02$. The sample with $x=0.0273$, instead,
displays a neat rise below 8~K, excluding the presence of regions with $%
x\simeq 0.02$. It can be concluded that the present measurements demonstrate
the presence of CSG domains also for $x<0.02$, and the insensitivity of the
temperature of the acoustic anomaly on doping supports the proposal\cite
{MFY01} of an intrinsic phase separation at $x<0.02${\em .}

\section{CONCLUSION}

We measured the anelastic spectra of La$_{2-x}$Sr$_{x}$CuO$_{4}$\ for both $%
x<0.02$\ and $x>0.02$, spanning the region of the phase diagram where the
transition from the SG to the CSG phase is expected. The same step-like
feature in the elastic energy loss function that has been attributed to the
domain wall motion in the CSG state is also found for $0.015<x<0.02$. The
step cannot be related to the SG phase whose onset is usually assumed to be $%
T_{f}\propto x$, since its temperature remains locked at the value found for 
$x=0.02$. The present data clearly show that CSG regions exist also for $%
x<0.02$ and therefore there is no neat separation between the SG and CSG
regions. They can be naturally interpreted in the framework of the phase
separation model for $x<0.02$, recently proposed by Matsuda {\it et al.}\cite
{MFY01} to explain their inelastic neutron scattering measurements.

\section*{Acknowledgments}

The authors thank Prof. A. Rigamonti for useful discussions. This work has
been done in the framework of the Advanced Research Project SPIS of INFM.

\begin{figure}[]
\caption{Derivatives of the logarithm of the Young's modulus with respect to
temperature at the HTT/LTO transformation for the five samples here studied.
The inset reports the nominal doping and the doping level deduced from the
temperature of the transition.}
\label{fig1}
\end{figure}

\begin{figure}[]
\caption{Elastic energy loss coefficient of La$_{2-x}$Sr$_{x}$CuO$_{4}$ with 
$x=0.0173$ ($1.35$~kHz), $x=0.0192$ ($1.05$~kHz), $x=0.0198$ ($0.76$~kHz) in
the part (a) of the figure and with $x=0.0227$ ($0.97$~kHz) and $x=0.0313$ ($%
1.72$~kHz) in part (b).}
\label{fig2}
\end{figure}

\begin{figure}[]
\caption{Derivative of the elastic energy loss with respect to temperature
for the five samples here studied.}
\label{fig3}
\end{figure}


\begin{references}
\bibitem{RBC98}  A. Rigamonti, F. Borsa and P. Carretta, Rep. Prog. Phys. 
{\bf 61}, 1367 (1998).

\bibitem{VSM87}  D. Vaknin, S.K. Sinha, D.E. Moncton, D.C. Johnston, J.M.
Newsam, C.R. Safinya and H.E.King,Jr. H. E. King, Jr., Phys. Rev. Lett. {\bf %
58}, 2802 (1987).

\bibitem{CCJ93}  J.H. Cho, F.C. Chou and D.C. Johnston, Phys. Rev. Lett. 
{\bf 70}, 222 (1993).

\bibitem{CBC93}  F.C. Chou, F. Borsa, J.H. Cho, D.C. Johnston, A.
Lascialfari, D.R. Torgeson and J. Ziolo, Phys. Rev. Lett. {\bf 71}, 2323
(1993).

\bibitem{CBJ92}  J.H. Cho, F. Borsa, D.C. Johnston and D.R. Torgeson, Phys.
Rev. B {\bf 46}, 3179 (1992).

\bibitem{JBC99}  M.-H. Julien, F. Borsa, P. Carretta, M. Horvatic, C.
Berthier and C.T. Lin, Phys. Rev. Lett. {\bf 83}, 604 (1999).

\bibitem{NBB98}  Ch. Niedermayer, C. Bernhard, T. Blasius, A. Golnik, A.
Moodenbaugh and J.I. Budnick, Phys. Rev. Lett. {\bf 80}, 3843 (1998).

\bibitem{CBK95}  F.C. Chou, N.R. Belk, M.A. Kastner, R.J. Birgeneau and A.
Aharony, Phys. Rev. Lett. {\bf 75}, 2204 (1995).

\bibitem{WUE99}  S. Wakimoto, S. Ueki, Y. Endoh and K. Yamada, Phys. Rev. B 
{\bf 62}, 3547 (2000).

\bibitem{CPC00}  F. Cordero, A. Paolone, R. Cantelli and M. Ferretti, Phys.
Rev. B {\bf 62}, 5309 (2000).

\bibitem{92}  R.S. Markiewicz, F. Cordero, A. Paolone and R. Cantelli, Phys.
Rev. B {\bf 64}, 54409 (2001).

\bibitem{MFY01}  M. Matsuda, M. Fujita, K. Yamada, R.J. Birgeneau, Y. Endoh
and G. Shirane, cond-mat/0111228 (2001).

\bibitem{DFF94}  M. Daturi, M. Ferretti and E.A. Franceschi, Physica C {\bf %
235-240}, 347 (1994).

\bibitem{NB}  A.S. Nowick and B.S. Berry, {\it Anelastic Relaxation in
Crystalline Solids}. (Academic Press, New York, 1972).

\bibitem{Joh97}  D.C. Johnston, {\it Handbook of Magnetic Materials}. ed. by
K.H.J. Buschow, p. 1 (North Holland, 1997).

\bibitem{SMM}  J.L. Sarrao, D. Mandrus, A. Migliori, Z. Fisk, I. Tanaka, H.
Kojima, P.C. Canfield and P.D. Kodali, Phys. Rev. B {\bf 50}, 13125 (1994).

\bibitem{LLN90}  W.-K. Lee, M. Lew and A.S. Nowick, Phys. Rev. B {\bf 41},
149 (1990).

\bibitem{61}  F. Cordero, C.R. Grandini, G. Cannelli, R. Cantelli, F.
Trequattrini and M. Ferretti, Phys. Rev. B {\bf 57}, 8580 (1998).

\bibitem{76}  F. Cordero, R. Cantelli and M. Ferretti, Phys. Rev. B {\bf 61}%
, 9775 (2000).

\bibitem{CPC01}  F. Cordero, A. Paolone, R. Cantelli and M. Ferretti, Phys.
Rev. B {\bf 64}, 132501 (2001).

\bibitem{MCP01}  R.S. Markiewicz, F. Cordero, A. Paolone and R. Cantelli,
Phys. Rev. B {\bf 64}, 54409 (2001).

\bibitem{GSB97}  R.J. Gooding, N.M. Salem, R.J. Birgeneau and F.C. Chou,
Phys. Rev. B {\bf 55}, 6360 (1997).

\bibitem{EPJB}  A. Campana , R. Cantelli , F. Cordero , M. Corti , A.
Rigamonti, E. Phys. J. B {\bf 18}, 49 (2000).
\end{references}
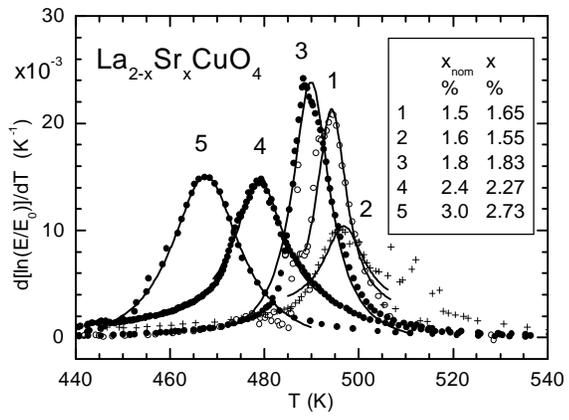
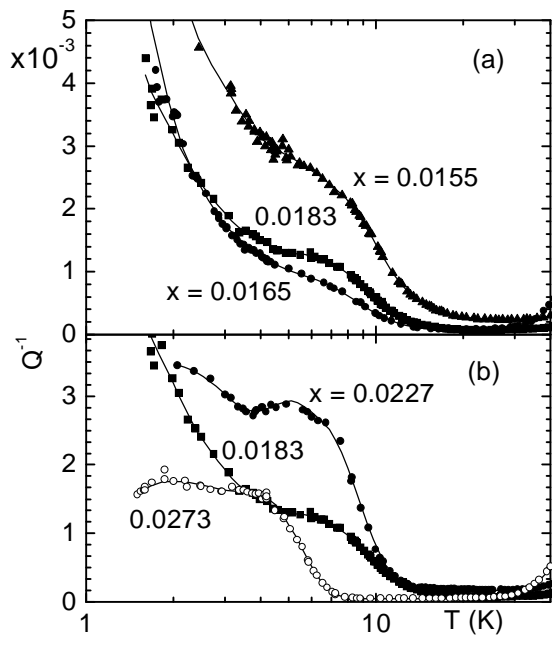
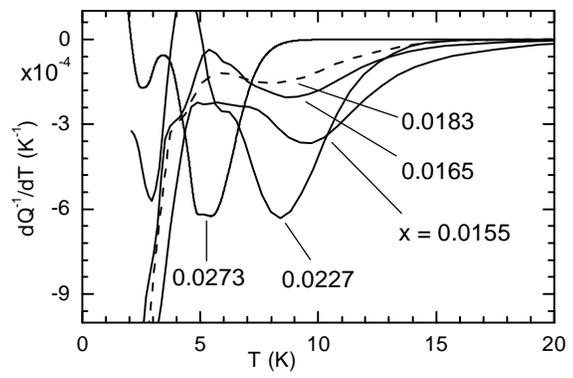
\end{document}